\documentstyle{fbssuppl}

\input epsf

\title{Single\ Meson\ Photoproduction\ via\ Higher\ Twist\ Mechanism and IR
Renormalons}
\author{S. S. Agaev}
\institute{High Energy Physics Lab., Baku State University,
Z.Khalilov st. 23,\\ 370148 Baku, Azerbaijan}

\begin{document}

\maketitle

\begin{abstract}
Single pseudoscalar and vector mesons hard semi-inclusive photoproduction $\gamma h\rightarrow
MX $ via higher twist (HT) mechanism is calculated using the QCD running coupling
constant method. The structure of infrared renormalon singularities of the HT 
subprocess cross section and the Borel sum for it
are found. The problem of normalization of HT process cross section in terms
of the meson elm form factor is discussed.
\end{abstract}

One of the fundamental achievements of QCD is the prediction of asymptotic
scaling laws for large-angle exclusive processes and their calculation in
the framework of pQCD [1-3]. In the context of the
factorized QCD an expression for an amplitude of an exclusive process can be
written as integral over ${\bf x}, {\bf y}$ of hadron wave functions (w.f.)\footnote{%
Strictly speaking, $\Phi _{M}({\bf x},Q^{2})$ is a
hadron distribution amplitude and it differs from a hadron wave function.
But in this paper we use these two terms on the same footing.} $\Phi
_{i}({\bf x},Q^{2})$ (an initial hadron), $\Phi_{f}^{*}({\bf y},Q^{2})$
(a final hadron) and amplitude $T_{H}({\bf x},{\bf y};\alpha _{S}({\hat Q}^{2}),Q^{2})$
of the hard-scattering subprocess [2].
This approach can be also applied for calculation of HT corrections to 
inclusive processes. The HT corrections
to a single meson photoproduction was studied in [4], 
where for computation of integrals over
${\bf x}, {\bf y}$, the frozen coupling approximation was used. In our work we 
consider the hard semi-inclusive photoproduction of single pseudoscalar and 
vector mesons $\gamma h\rightarrow MX$ using the running coupling constant method.

The two HT subprocesses, namely $\gamma q_{1}\rightarrow Mq_{2}$
and $\gamma \overline{q}_{2}\rightarrow M\overline{q}_{1}$ contribute to the
photoproduction of the single meson $M$ in the reaction $\gamma h\rightarrow
MX$ . 
The amplitude for the subprocess $\gamma q_{1}\rightarrow Mq_{2}$ can be
found by means of the Brodsky-Lepage method [2],
\begin{equation}
M=\int_{0}^{1}\int_{0}^{1}dx_{1}dx_{2}\delta
(1-x_{1}-x_{2})T_{H}(x_{1},x_{2};\alpha _{S}(\hat{Q}^{2}),~\hat{s},
~\hat{u},~\hat{t})\Phi _{M}(x_{1},x_{2};Q^{2})  \label{2}
\end{equation}

In Eq.(1) $\Phi _{M}$ is the meson $M$ w.f. In this work we use the following
w.f.;
for the pion and ${\rho}$-meson  
\begin{equation}
\Phi _{M}(x,\mu _{0}^{2})=\Phi _{asy}^{M}(x)\left[ a+b(2x-1)^{2}\right] .\label{3}
\end{equation}
where, $a=0,b=5$ (pion), $a=0.7,b=1.5$ (longitudinally and transversely 
polarized ${\rho}$-meson) and 
for the kaon
\begin{eqnarray}
\Phi _{K}(x,\mu _{0}^{2}) &=&\Phi _{asy}^{K}(x)\left[
a+b(2x-1)^{2}+c(2x-1)^{3}\right] ,  \label{5} \\
a &=&0.4,~~b=3,~~c=1.25.  \nonumber
\end{eqnarray}
with $\Phi _{asy}^{M}(x)=\sqrt{3}f_{M}x(1-x)$ being the meson M asymptotic w.f.
The mesons' decay constants $f_{M}$ take values $f_{\pi}=0.093$ GeV,
$f_{K}=0.112$ GeV, $f_{\rho}^{L}=0.2$ GeV, $f_{\rho}^{T}=0.16$ GeV.

The details of calculation of the HT subprocess cross section
are described in our work [5], where 
expressions for $d\hat{\sigma }%
^{HT}/d\hat{t}$ can be found. The subprocess cross section depends 
on quantities $I_{1,2}$, $K_{1,2}$,

\begin{equation}
I_{1}(K_1)=\int_{0}^{1}\int_{0}^{1}\frac{dx_{1}dx_{2}\delta (1-x_{1}-x_{2})\alpha
_{S}(\hat{Q}_{1}^{2}(\hat{Q}_{2}^{2}))\Phi _{M}(x_{1},x_{2})}{x_{2}(x_1)},
\label{12}
\end{equation}

\begin{equation}
I_{2}(K_2)=\int_{0}^{1}\int_{0}^{1}\frac{dx_{1}dx_{2}\delta (1-x_{1}-x_{2})\alpha
_{S}(\hat{Q}_{1}^{2}(\hat{Q}_{2}^{2}))\Phi _{M}(x_{1},x_{2})}{
x_{1}x_{2}},  \label{13}
\end{equation}
where for $I_{1},I_{2}$ the renormalization and factorization scale is $%
\hat{Q}_{1}^{2}=x_{2}\hat{s}$, for $K_{1},K_{2}$ it is given by $%
\hat{Q}_{2}^{2}=-x_{1}\hat{u}$.

In the frozen coupling approximation
one puts $\hat{Q}_{1,2}^{2}$ equal to their mean values $\hat s/2, -\hat u/2$ and
removes $\alpha_{S}(\hat{Q}_{1,2}^{2})$ as the constant factor in Eqs(4-5). After such
manipulation integrals (4-5) are trivial and can be easily computed. In this approach
the single meson photoproduction cross section can be normalized in terms of
the meson elm form factor only if the meson w.f. is symmetric under
replacement $2x-1 \leftrightarrows 1-2x$ (pion, ${\rho}$-meson).

In the framework of the running coupling method, for example, $I_{1}$ takes 
the form
\begin{equation}
I_{1}(\hat{s})=\int_{0}^{1}\frac{\alpha _{S}((1-x)\hat{s})\Phi
_{M}(x,\mu _{0}^{2})dx}{1-x}.  \label{21}
\end{equation}
The $\alpha _{S}((1-x)\hat{s})$ has the infrared singularity at $%
x\rightarrow 1$ and as a result integral (6) diverges. This divergence
is induced by ir renormalons. Indeed, the Borel transform $B[I_1](u)$ of
$I_1(\hat {s})$ has ir renormalon poles at $u=1,2,3,4$ (for w.f. (2)) [5].
The integral (6) can be regularized by means of the principal value
prescription. The Borel sum (resummed expression) of $I_{1}$ is
\begin{eqnarray}
\left[ I_{1}\left( \hat{s}\right) \right] ^{res}&=&\frac{4\sqrt{3}\pi f_{M}%
}{\beta _{0}}\left[ \left( a+b\right) \frac{Li(\lambda )}{\lambda }-\left(
a+5b\right) \frac{Li(\lambda ^{2})}{\lambda ^{2}}\right.
\nonumber\\
&&\left. +8b\frac{Li(\lambda ^{3})}{%
\lambda ^{3}}-4b\frac{Li(\lambda ^{4})}{\lambda ^{4}}\right] ,  \label{30}
\end{eqnarray}
where $Li(\lambda )$ is the logarithmic integral, for $\lambda >1$
defined in its principal value
\begin{equation}
Li(\lambda )=P.V.\int_{0}^{\lambda}\frac{dx}{\ln x}%
,~~~~\lambda =\hat{s}/\Lambda ^{2}.  \label{31}
\end{equation}

The similar expressions can be found for $I_2$ and $K_{1,2}$.

Important question here is the normalization of the
meson photoproduction cross section in terms of the meson elm form
factor. The pion and kaon form factors have been calculated by means of the
running coupling method in our papers [6-7]. 
Using the expressions obtained in this works and our recent results,
it is not difficult to conclude that,
in the running coupling approach the
HT subprocess cross section cannot be normalized in
terms of the meson form factor neither for mesons with symmetric w.f. nor
for non-symmetric ones.

Some of our numerical results for the photon-proton process are plotted in Fig.1. 
Here, for calculation
of ratios $r_{M},~R_{M}$
the $\Sigma_{M^{+(-)}}=d\sigma(\gamma p \rightarrow M^{+(-)}X)$ 
inclusive cross sections and the difference $\Delta_{M}=\Sigma_{M^{+}}-\Sigma_{M^{-}}$ 
are used. In $\Sigma_{M^{+(-)}}$, $\Delta_M$ the dominant
leading twist LT ($\gamma q \rightarrow gq$ with $q\rightarrow M$)
and HT ($\gamma q\rightarrow Mq$) contributions to the photoproduction
have been taken into account. 

\begin{figure}[tb]
\epsfxsize=7cm
\epsfysize=7cm
\centerline{\epsffile{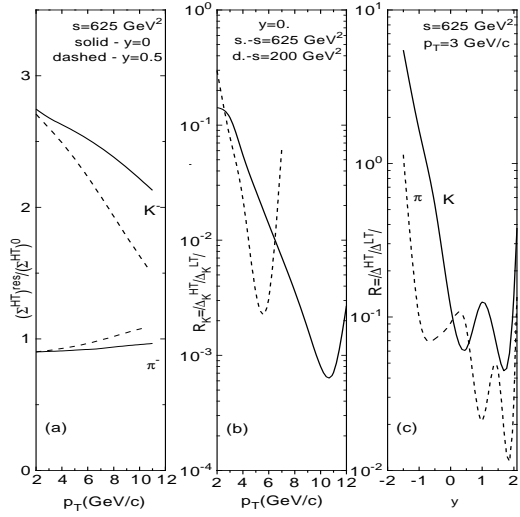}}
\caption{\label{fig:trent.ps} a) Ratio $r_M=(\Sigma_{M}^{HT})^{res}/(\Sigma_{M}^{HT})^{0}$,
where $\Sigma^{res,(0)}$ are HT contributions to the photoproduction cross section calculated
using the running and frozen coupling approximations, respectively;b) $R_M$ for
the kaon, as a function of $p_T$; c) $R$ for $\pi,~K$ as a function of y}
\end{figure}

As seen from Fig.1(a), effect of ir renormalons is considerable for
$K^{-}$, whereas for $\pi^{-}$ we have $r_{M}\simeq 1$.
In Fig.1(b) the ratio $R_{M}=\mid \Delta_{K}^{HT}/
\Delta_{K}^{LT} \mid$ is shown. For all particles the LT cross section 
difference is positive $\Delta_{M}^{LT}>0$, since 
$\Sigma_{M^{+}}^{LT}\sim u_{p}(x,-{\hat t})e_{u}^{2}$, while 
$\Sigma_{M^{-}}^{LT}\sim d_{p}(x,-{\hat t})e_{d}^{2}$. The smaller quark charge
$e_{d}$ and the smaller distribution function $d_{p}$ both suppress 
$\Sigma_{M^{-}}^{LT}$. The HT cross section difference may change
sign at small $p_{T}$ and become negative $\Delta_{M}^{HT}<0$. 
Therefore, we plot the absolute value of $R_{M}$. The similar picture 
has been also found for other mesons. 
The rapidity dependence of $R_{M}$ at $\sqrt{s}=25$ GeV, $p_{T}=3$ GeV/c plotted in
Fig.1(c) illustrates not only the tendency of the HT contributions to be
enhanced in the region of negative rapidity, but also reveals an interesting
feature of the HT terms; as is seen from Fig.1(c) the ratio $R_{M}$ is an
oscillating function of the rapidity. This property of the HT terms has
important phenomenological consequences in the case of $\rho$-meson
photoproduction; comprehensive analysis of these effects can be found in ref.[5].

Summing up we can state that:\\
i) for mesons with non-symmetric w.f. in the framework of the frozen coupling 
approximation the higher twist subprocess cross section cannot be normalized
in terms of a meson electromagnetic form factor;\\
ii) in the context of the running coupling constant method the HT subprocess
cross section cannot be normalized in terms of meson's elm form factor
neither for mesons with symmetric w.f. nor for non-symmetric ones;\\
iii) the resummed HT cross section differs from that found using the frozen
coupling approximation, in some cases, considerably;\\
iv) HT contributions to the single meson photoproduction cross section have
important phenomenological consequences, specially in the case of $\rho$-meson
photoproduction. In this process the HT contributions wash the LT results off,
qualitatively changing the LT predictions.

\end{document}